\documentclass[prb,twocolumn,secnumarabic,floatfix,amssymb,amsmath,showpacs,aps]{revtex4-1}

\usepackage{graphicx,color,epsfig,rotate}
\usepackage{times}
\usepackage{color}
\usepackage{graphicx}
\usepackage{dcolumn}
\usepackage{amssymb}
\usepackage{amsmath}
\usepackage{amsfonts}
\usepackage{bm}%
\usepackage{wrapfig}
\usepackage[colorlinks=true,linkcolor=blue,pagecolor=blue,filecolor=blue,menucolor=blue,urlcolor=blue,citecolor=blue,anchorcolor=blue]{hyperref}%
\usepackage{bm}
\usepackage{graphics}
\usepackage{graphicx}
\usepackage{color}
\usepackage{amssymb}
\usepackage{standalone}
\usepackage[toc,page]{appendix}
\usepackage{import}
\usepackage{enumerate}
\usepackage{sidecap}

\newcommand{\lpa}[1]{\left({#1}\right)}    
\newcommand{\bpa}[1]{\left[{#1}\right]}    
\newcommand{\lca}[1]{\lbrace{#1}\rbrace}   
\newcommand{\ket}[1]{\vert{#1}\rangle}     
\newcommand{\mel}[3]{\langle{#1}\vert {#2} \vert{#3}\rangle} 
\newcommand{\pro}[1]{\vert{#1}\rangle \langle{#1}\vert}  
\newcommand{\cop}[2]{\vert{#1}\rangle \langle{#2}\vert}  

\newcommand{\tl}{\tilde}                   
\newcommand{\mcl}[1]{\mathcal{#1}}         
\newcommand{\hcn}[1]{#1^{\dagger}}
\newcommand{\beq}{\begin{equation}}                        
\newcommand{\eeq}{\end{equation}}                          

\newcommand{\pderiv}[2]{\frac{\partial #1}{\partial #2}}

\begin{document}

\title{Indirect exciton qubit manipulation via the optical Stark effect in quantum dot molecules}

\author{J. E. Rolon}
\email{rolon@email.unc.edu}
\affiliation{Department of Physics and Astronomy, University of North Carolina, Chapel Hill, North Carolina, 27599-3255, USA}

\author{J. E. Drut}
\email{drut@email.unc.edu}
\affiliation{Department of Physics and Astronomy, University of North Carolina, Chapel Hill, North Carolina, 27599-3255, USA}

\date{\today}

\begin{abstract}
We propose a coherent control scheme based on the optical Stark effect in optically generated excitons in quantum dot molecules (QDMs). We show that, by
the combined action of voltage bias detuning sweeps and Rosen-Zener pulsed interactions, it is possible to dynamically generate and modify an anticrossing
gap that emerges between the dressed energy levels of long-lived, spatially indirect excitons. We perform numerical and analytic non-perturbative calculations
based on the Bloch-Feshbach formalism, which demonstrate that this effect induces a mechanism of coherent population trapping of indirect excitons in QDMs.
Our results show that it is possible to perform an all-optical implementation of indirect-excitonic qubit operations, such as the Pauli-X and Hadamard quantum
gates, across two defined axis of the Bloch Sphere.
\end{abstract}

\maketitle

\section{Introduction}
\label{SectionIntro}

Quantum dot molecules (QDMs) are promising building blocks for semiconductor-based approaches to scalable quantum information technologies~\cite{QDMBook}. They possess remarkable properties, among which are long-lived charge and spin states of confined carriers and excitons~\cite{Krenner, Stinaff1}, tunable exciton relaxation rates~\cite{Stinaff2}, sustained coherent Rabi oscillations~\cite{VillasBoas, Ardelt}, and the ability to couple spin and charge to photonic cavity modes~\cite{GammonCavity}. Moreover, the tunability of the QDM exciton spectrum, and the selective excitation of the spin and charge degrees of freedom, make exciton-based coherent control protocols a viable route for the implementation of universal quantum gates~\cite{ExcitonicQuantumInfo}. In this context, having controllable exciton states that are resilient against the effects of decoherence is a fundamental requirement for the construction of the corresponding qubits. In particular, spatially indirect (neutral) excitons in QDMs have been proposed as suitable excitonic qubits owing to their extended lifetimes and robust electrically-controlled optical properties~\cite{RolonControl, RolonItaly}, and are the object of active experimental and theoretical research efforts for the engineering of quantum information schemes in QDMs~\cite{Stinaff2, RolonStinaff, Alcalde1, Alcalde2}.

Apart from the purely electric (or magnetic) means of controlling exciton states, ultrafast optical excitation can induce profound modifications to the excitonic level structure of single quantum dots (QDs) and QDMs. A prominent example is the optical Stark effect~\cite{StarkEffectBook}, which is manifested as a quasi-static shift of exciton line shapes when the system is subjected to intense ultrafast laser excitation pulses. In QDs, transient reflectivity measurements with pulsed non-resonant excitation display a spectral envelope that depends on the strength of the excitation field~\cite{Unold}. Therefore, the combined action of the optical Stark effect and the application of external electric fields could serve as a useful mechanism for the implementation of coherent control of spin and exciton states in QDs and QDMs~\cite{Ramsay, Brash, Imamoglu, Xu, Nazir}.

In this work, we study the influence of the optical Stark effect on the spectra and dynamics of indirect excitonic qubits in electrically gated and optically driven QDMs~\cite{RolonControl}. We propose that, through the combined action of ultrafast laser pumping with a time-dependent Rosen-Zener (RZ) pulse envelope~\cite{RosenZener} and the application of a bias detuning of the indirect energy levels, it is possible to dynamically open and close the gap of a cotunneling-induced anticrossing between spatially indirect excitons. We demonstrate that this mechanism enables the coherent population trapping of either of the avoided crossed excitons, and further allows coherent control of indirect excitonic qubits about two axis of the Bloch sphere.

This paper is organized as follows. Section \ref{SectionModel} introduces a realistic phenomenological model of the QDM exciton level structure, which takes into account the parametric dependence of single charge confinement energies and interdot couplings on the structural parameters of the system. We also discuss the numerical methods used to compute the population dynamics of the exciton states. In Sec.\ \ref{LACSPopulationSpectrum}, we present the reconstruction of the QDM exciton level anticrossing spectrum (LACS) using level population bias maps. These maps help us identify the different molecular resonances and optical signatures as a function of the applied bias voltage, laser excitation frequency, and intensity. Section \ref{SectionDynamicStark} discusses the role of the optical Stark effect on the spectral characteristics and dynamics of a qubit manifold comprised by spatially indirect excitons. By means of a non-perturbative calculation, we present comprehensive analytical expressions that show the dependence of the indirect-excitonic dressed energies, avoided crossings, and interactions, on the bias detuning and the matrix elements that couple the qubit manifold to the driving fields.
In Sec.\ \ref{SectionCoherentControl}, we discuss the behavior of the exciton population dynamics resulting from the implementation of bias detuning and RZ pulses which modulate the intensity of the optical Stark shift. We show that optical Stark effect generates a coherent population trapping mechanism which can be used to perform indirect-excitonic qubit rotations about two axis of the Bloch sphere.
\begin{figure}[t]
\includegraphics[width=0.7\columnwidth]{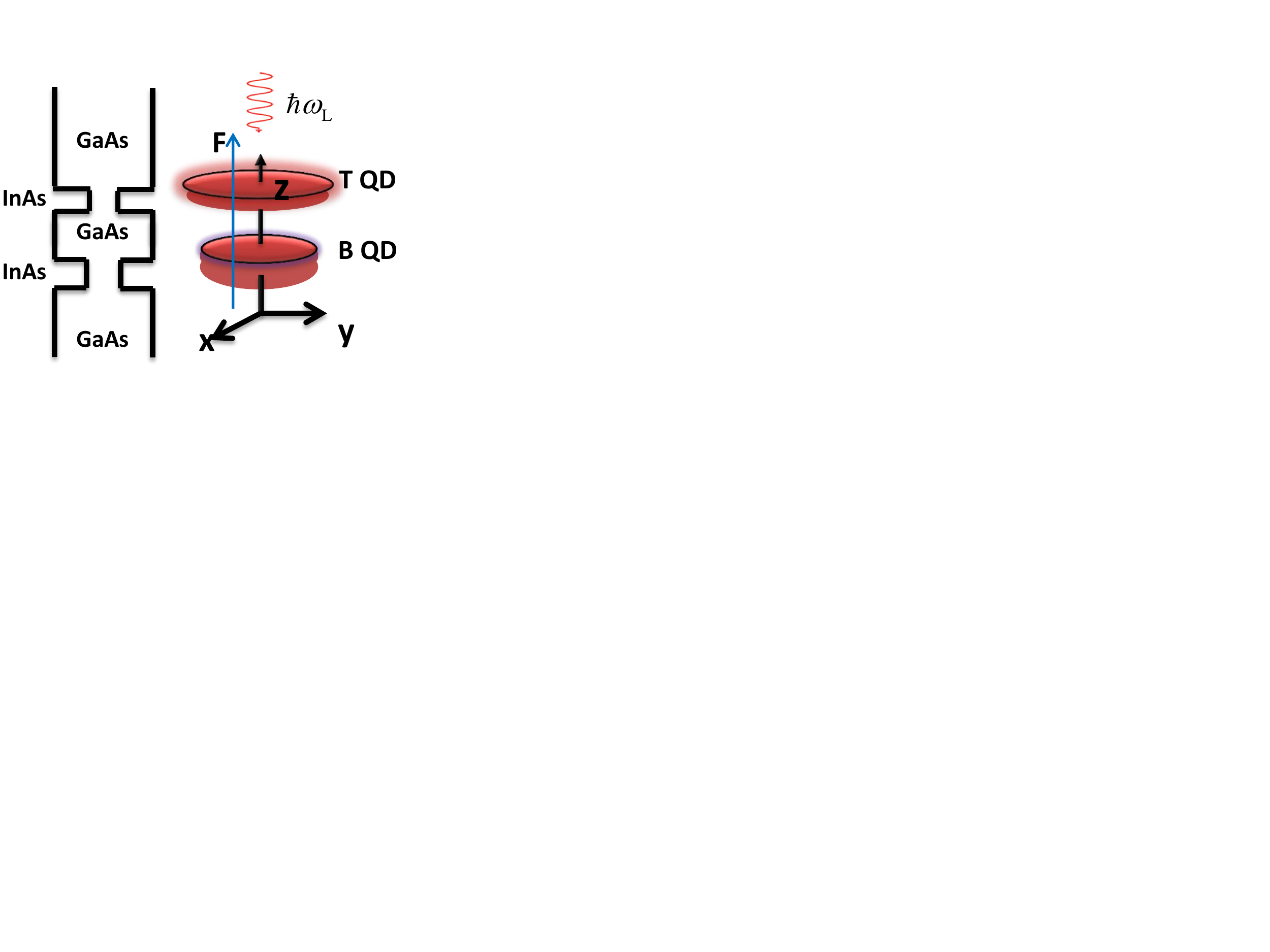}
\caption{(Color online) Schematic representation of the quantum dot molecule discussed in this work. See Sec.~\ref{SectionModel} for details.}
\label{Fig:system}
\end{figure}

\section{Model}
\label{SectionModel}

The system under consideration consists of two vertically stacked QDs embedded in a $n$-$i$ Schottky junction photodiode (see Fig.~\ref{Fig:system}), as typically used in photoluminescence and pump-probe spectroscopy experiments~\cite{Krenner, FinleyPumpProbe}. The QDs are separated by a barrier of thickness $d$ and subjected to an applied axial electric field $F$. In our model, single neutral excitons are pumped by a broad square laser pulse of frequency, $\omega_L$, with a time-dependent electric field envelope $E(t)$. The excitonic bare states are denoted according to single charge occupation in each QD, i.e. $_{h_{B}h_{T}}^{e_{B}e_{T}}X$, where $e_{B(T)},h_{B(T)}=\lbrace 0,1\rbrace$ are the electron and hole occupation numbers in the ``bottom (B)" and ``top" (T) QDs, respectively.
These states comprise the following basis~\cite{BiexComment}: the vacuum state
\beq
\ket{1} \equiv \ket{_{00}^{00}X};
\eeq
two neutral {\em spatially direct} exciton states:
\beq
\ket{2} \equiv \ket{_{10}^{10}X} \ \ \ \ \text{and} \ \ \ \ \ket{5} \equiv \ket{_{01}^{01}X};
\eeq
and two neutral {\em spatially indirect} exciton states:
\beq
\ket{3} \equiv \ket{_{10}^{01}X} \ \ \ \ \text{and} \ \ \ \ \ket{4} \equiv \ket{_{01}^{10}X}.
\eeq
In this basis, the Hamiltonian is
\begin{equation}
\begin{split}
\mcl{H} &= \sum_{i=1}^{5} E_i \pro{i} + \sum_{j=2,5} (\hbar\Omega_{j}(t) e^{-i\omega_L t} \cop{1}{j}) +V_F \cop{2}{5} \\
  &+ t_e (\cop{2}{3} + \cop{5}{4}) + t_h(\cop{2}{4} + \cop{3}{5}) + \text{H.c.} \, ,
\end{split}
\label{FullHamiltonian}
\end{equation}
where the $E_i$'s represent the exciton bare energies~\cite{PawelX, Pawel, UsmanQuantitative, Bester} and $t_e, t_h$ the electron and hole tunneling matrix elements~\cite{tunnelings}. Here, $V_F=\frac{{\mu}_B \, {\mu}_T}{4\pi\epsilon_0\epsilon_{r}d^{3}}$ accounts for interdot exciton hopping processes mediated by dipole-dipole interactions~\cite{FRETRolon, FRETGeneral}, with $\epsilon_{r}$ being the dielectric constant, and $\mu_{T(B)}$ the interband transition dipole moments~\cite{dipoles1, dipoles2, dipoles3}.

\begin{figure}[h]
\includegraphics[width=1.0 \columnwidth]{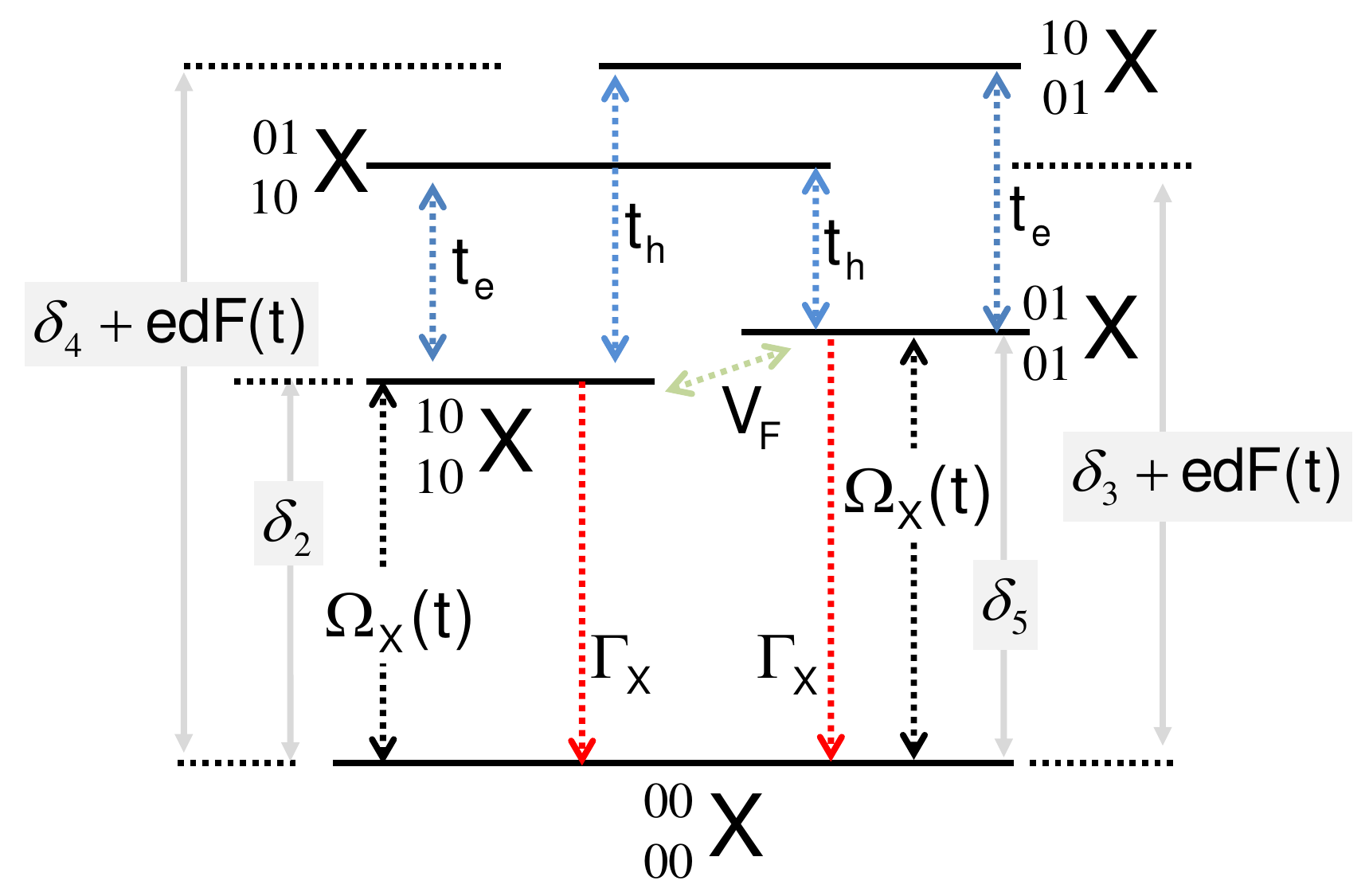}
\caption{(Color online) Single exciton level configuration diagram. Dashed arrows indicate couplings mediated by different processes: electron and hole tunneling, $t_e,t_h$; optical coupling, $\Omega_X(t)$, and exciton hopping, $V_F$. Relaxation channels are indicated by red dashed arrows, while solid arrows in gray indicate the respective laser-exciton transition energy detunings $\delta_i=E_i-\hbar\omega_L$.}
\label{LevelConfigDiagFig}
\end{figure}

The Hamiltonian dynamics is controlled via two parameters. The first one is achieved via pulsing the applied bias voltage $F(t)$, which drives the detuning of the spatially indirect transition energies,
\begin{eqnarray}
E_3 &\rightarrow& E_3 - edF(t),\\
E_4 &\rightarrow& E_4 + edF(t).
\end{eqnarray}
The second parameter is achieved through the optical Stark effect resulting from pulsing the shape of the laser electric field envelope, i.e. the time-dependence of the exciton coupling to the laser field given by
\beq
\Omega_j(t) = \mel{1}{\vec{\mu}_j \cdot \vec{E}(t)}{j}.
\eeq
The applicability of our model does not rely on any resonant condition between the bare exciton energy levels at zero bias. The simulation parameters utilized in our model are given in Ref.\ [\onlinecite{parameters}] and a level configuration diagram of the Hamiltonian is shown in Fig.\ \ref{LevelConfigDiagFig}.

In order to extract the exciton dynamics, molecular resonances, and the response of the system to the effect of the control pulses, we solve for the density matrix, whose time evolution is governed by the Markovian master equation
\begin{equation}
\pderiv{\rho}{t} = -\frac{i}{\hbar}\bpa{\mcl{H},\rho} + \mcl{L}\rho \, .\label{Lindblad}
\end{equation}
In our formalism, the Liouvillian $\mcl{L}\rho$ has the Lindblad form given by
\begin{equation}
\mcl{L}\rho = \sum_i \frac{1}{2} \Gamma_i \lpa{2\alpha_i\rho\alpha_i^{\dagger} -\alpha_i^{\dagger}\alpha_i\rho + \rho\alpha_i^{\dagger}\alpha_i}\,
\end{equation}
where $\hcn{\alpha}_i$ ($\alpha_i$) and $\Gamma_i \simeq 1\textrm{ns}^{-1}$ are the exciton creation (annihilation) operators and effective relaxation rates for channel $i$, respectively. For appropriately chosen QDM structural parameters and excitation conditions, the characteristic periods $\tau_{c}$ (corresponding to coherent oscillations of the exciton populations $\rho_{ii}(t)$) could be much shorter than the exciton recombination time, i.e. $\tau_c \ll\tau_X $; therefore, the exciton population dynamics is nearly coherent for times $\tau_c < t \ll \tau_{X}$ \cite{lifetimes1,lifetimes2, coherent-note-1}.

\section{Level anticrossing spectrum}
\label{LACSPopulationSpectrum}

In order to obtain the behavior of the different exciton molecular resonances, and its dependence on the coupling parameters, applied electric field and excitation power, we construct a level anticrossing map. We do this in two different but equivalent ways, as we explain next.

First, by diagonalization of the Hamiltonian Eq.\ \ref{FullHamiltonian} in the rotating wave approximation (RWA)~\cite{RWA}, for fixed excitation energy $\hbar\omega_L$, while varying only the bias voltage $F$. We thus obtain a global picture of the energy eigenvalues (dressed energies), eigenvector components, and the electrically-tunable spatial character of the molecular excitons, which is governed mainly by charge tunneling and the optical Stark effect.

In the second approach, we reconstruct an averaged level occupation map corresponding to each individual exciton state by direct integration of the diagonal elements of the density matrix,
\beq
p_i = \frac{1}{t_{L}}\int_{0}^{t_{L}}\rho_{ii}(t)dt,
\eeq
obtained by solutions to Eq.\ \ref{Lindblad}. Here, $t_L$ stands for a constant-amplitude pulse duration that is long enough to capture several Rabi oscillations of the exciton populations; in practice, a few Rabi oscillations are enough to reliably compute $p_i$. Therefore, each exciton pumped into the system will exhibit a relative amplitude $p_i(F, \hbar\omega_L, \Omega_X)$.

Alternatively, one can compute the level occupation map of the vacuum state (RWA photon field) $\ket{1} \equiv \ket{_{00}^{00}X}$, so that the complete dressed LACS spectrum will be mapped by all coordinates $(F, \hbar\omega_L, \Omega_X)$ where this state is {\em depopulated}, i.e. by transferring its population to each corresponding exciton. In particular, the latter approach enables us to estimate the excitation power dependence of the different spectral signatures resembling those obtained by photoluminiscence or pump-probe spectroscopy.

\begin{figure}[h]
\includegraphics[width=1.0 \columnwidth]{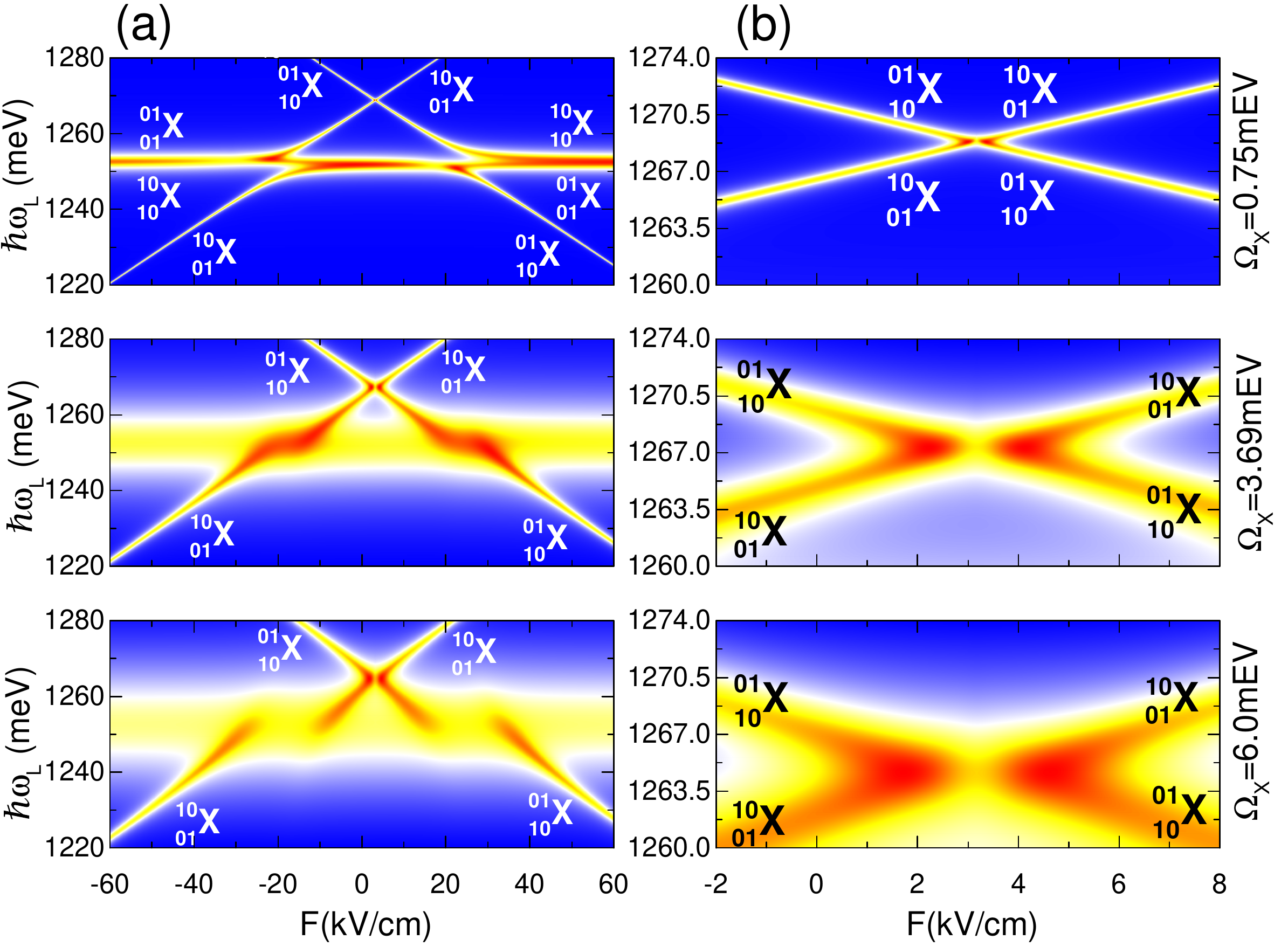}
\caption{(Color online) Level anticrossing map of $|_{00}^{00}X\rangle$ as function of the applied electric field, $F$, and laser excitation energy, $E_{\text{laser}}=\hbar\omega_L$, and for different values of the optical coupling, $\Omega_X$. (a) Full spectrum showing electron tunneling anticrossing signatures at $\hbar\omega_L \simeq 1252$meV, $F\simeq (-18.7, 23.4)$kV/cm, together with the asymptotic behavior of molecular sates towards the spatially direct, $|_{10}^{10}X\rangle$, $|_{01}^{01}X\rangle$ and indirect exciton states, $|_{10}^{01}X\rangle$, $|_{01}^{10}X\rangle$. (b) The spatially indirect exciton manifold shows a robust anticrossing signature at $F\simeq 3.17$kV/cm, which persists under strong power broadening effects caused by large values of $\Omega_X$.}
\label{LevelAnticrossingMapFig}
\end{figure}

Figure \ref{LevelAnticrossingMapFig} (a) shows the level anticrossing spectrum corresponding to the energy level diagram in Fig.\ \ref{LevelConfigDiagFig}. This was reconstructed from the level occupation map of the vacuum state $|_{00}^{00}X\rangle$ as a function of applied electric field $F$ and laser excitation energy $\hbar\omega_L$, for three values of the optical coupling strength $\Omega_X = 0.75, 3.69, 6.0$meV. The spectral pattern shows anticrossing signatures between spatially direct and indirect excitonic molecular states at $F\simeq -18.7$ and $F\simeq 23.4$kV/cm, each having a gap of $\simeq 6.23$meV; these are mainly the result of electron tunneling. Clearly, with increasing optical coupling, the direct states $|_{10}^{10}X\rangle$, $|_{01}^{01}X\rangle$ become prominently affected by power broadening effects such that, for the highest value of the coupling, the electron tunneling anticrossing intersecting the horizontal line shape $\hbar\omega_L \simeq 1252$meV becomes almost entirely ``washed out". In contrast, the spatially indirect exciton spectral lines, $|_{10}^{01}X\rangle$ and $|_{01}^{10}X\rangle$, are more robust to the effects of increasing $\Omega_X$, as their broadening weakens as $\vert F \vert$ is increased respect to the position of the tunneling resonances. This can be easily understood: the optical coupling $\Omega_I$ of the indirect excitons is mediated by the optical pumping of electrons and holes and their tunneling rates, which become weaker with increasing $\vert F \vert$ away from the avoided crossings.

On the other hand, the optical signature in Fig.\ \ref{LevelAnticrossingMapFig}(b) shows a robust anticrossing at $F\simeq 3.17$kV/cm ($\hbar\omega_L \simeq 1269$meV) reflecting the mixing of the two indirect states $|_{10}^{01}X\rangle$ and $|_{01}^{10}X\rangle$, which persists even under strong power broadening effects caused by large values of $\Omega_X$ (the signature shifts to lower energies as a result of the optical Stark effect, as explained in Fig.\ \ref{LevelEigenspectrumFig}). Indeed, as shown in our previous work, see Ref.\ [\onlinecite{RolonControl}], a spectrally isolated two-level system spanned by the indirect excitons shows an anticrossing signature which can form the basis for constructing {\it indirect excitonic qubits} resilient to spontaneous recombination, and whose coherent dynamics at low excitation powers ($\Omega_X \ll 1\text{meV}$) can be controlled by the tuning the electric field $F$. However, as shown here, by means of the optical Stark effect it is possible to take advantage of laser-pulse shaping techniques to control the coherent dynamics of the spatially indirect excitons from low to moderately high excitation powers.

\section{The optical Stark effect}
\label{SectionDynamicStark}

\begin{figure}[h]
\includegraphics[width=1.0 \columnwidth]{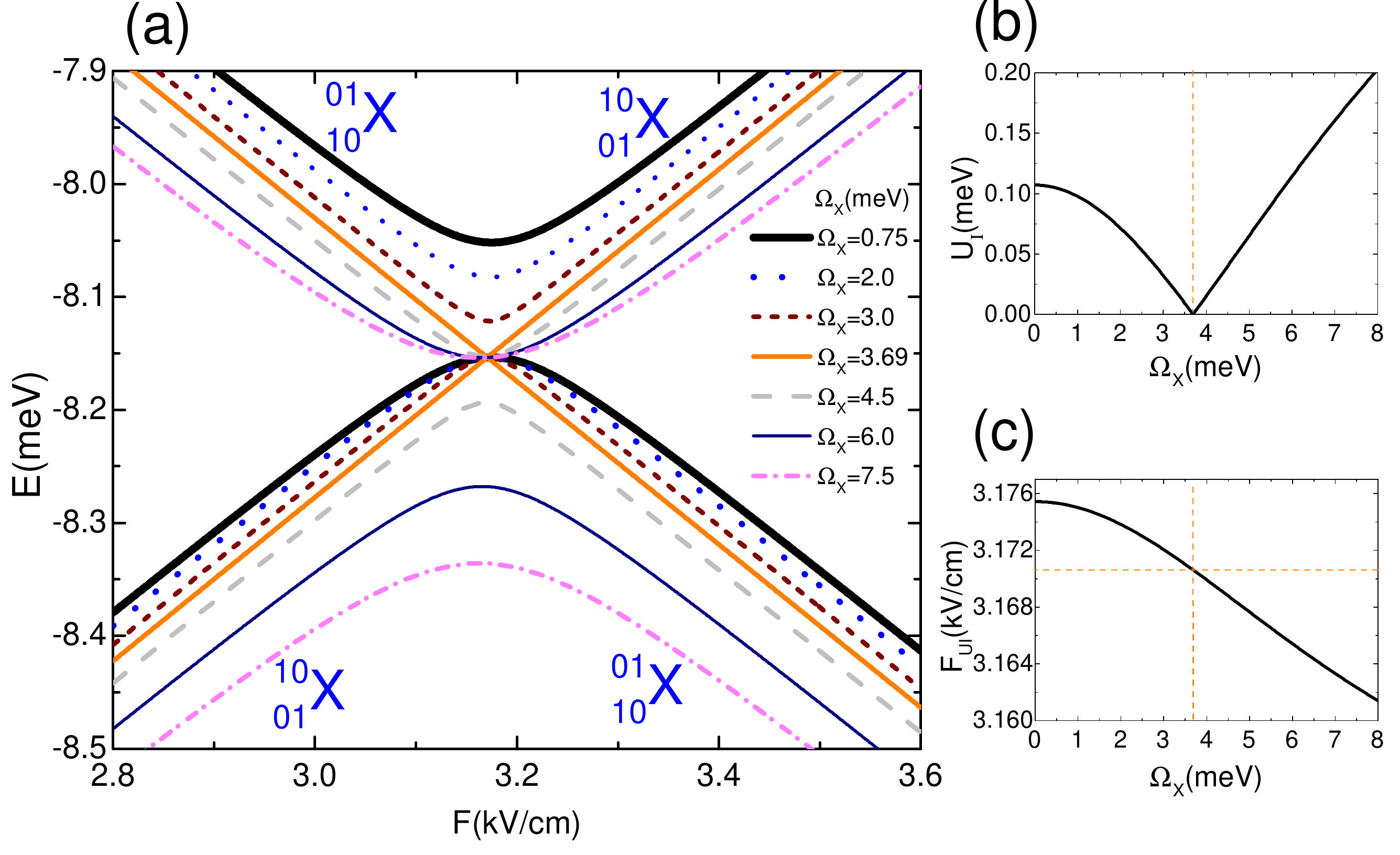}
\caption{(Color online)(a) Bias-dependent exciton dressed energy spectrum at fixed excitation energy, $\hbar\omega_L=1277$meV, showing the anticrossing of the indirect exciton states for different values of the optical coupling $\Omega_X$. As a result of the optical Stark shift, the anticrossing gap vanishes for $\Omega_X^{(c)} \simeq 3.69$meV (solid line in orange) (b) Anticrossing gap minima as function of $\Omega_X$ (including a correction for the shifting of the minima along the bias voltage axis, $F$, as shown in (c).}
\label{LevelEigenspectrumFig}
\end{figure}

In order to elucidate the origin and behavior of the optical Stark effect signatures shown in Fig.\ \ref{LevelAnticrossingMapFig}(b), for different values of the optical coupling strength $\Omega_X$, we first calculate numerically the excitonic dressed energy spectrum as function of the applied electric field $F$, for a fixed non-resonant excitation energy $\hbar\omega_L=1277$meV ($\hbar\omega_L$ is detuned from both the direct and indirect exciton transition energies). Figure \ref{LevelEigenspectrumFig}(a) shows the anticrossing signature resulting from the mixing of the spatially indirect states $|_{10}^{01}X\rangle$ and $|_{01}^{10}X\rangle$. Remarkably, as the optical coupling increases from $\Omega_X = 0.75$meV (solid line in black) the gap shows a non-monotonic behavior, vanishing for $\Omega_X^{(c)} \simeq 3.69$meV (solid line in orange). Figure \ref{LevelEigenspectrumFig} shows the value of the anticrossing gap as function of $\Omega_X$, with the position of the minimum indicated by the red dashed line. As shown in Fig.\ \ref{LevelEigenspectrumFig}(c), the position of the gap minima along the bias detuning axis slightly shifts as we vary the optical coupling, reaching a critical value, $F_c$, when the gap vanishes. The observed shift of the gap minima is a consequence of the of the parametrical dependence of the dressed exciton energies on the values of the interdot couplings and energy detunings of the indirect exciton transitions.

In the following subsections, by means of a non-perturbative calculation, we present comprehensive analytic expressions which reveal the origin of the aforementioned effect in terms of the interplay between the energy and intensity of the driving field, and all of the relevant interdot couplings and energy detunings. We also discuss how the tunability of the indirect excitonic gap via the optical Stark effect enables a dynamically controlled coherent population trapping mechanism, with potential applications to the coherent control of indirect excitonic qubits in QDMs.

\subsection{Non-perturbative calculation of the optical Stark effect for indirect excitons}
\label{SubsectionNonPerturbativeCalc}

As discussed above, the appearance of a level anticrossing signature between spatially indirect excitons points to the onset of a non-trivial quantum coherent interaction. Moreover, the dependence of this interaction on the optical coupling $\Omega_X$ cannot be straightforwardly explained by the off-diagonal matrix elements of the Hamiltonian in Eq.\ \ref{FullHamiltonian}, nor from the level diagram shown in Fig.\ \ref{LevelConfigDiagFig}. However, the physics can be revealed by an effective Hamiltonian $\mcl{H}_\text{eff}$ resulting from the projection of the full Hamiltonian onto a reduced sector of the Hilbert space containing eigenvectors relevant to the anticrossing region, with eigenvalues matching exactly those of the full Hamiltonian. In other words, we aim to adiabatically eliminate the spatially direct excitonic sector of the Hamiltonian, while retaining its dynamical effects by including (to all orders) the resulting perturbative corrections to the matrix elements of the projected Hamiltonian. To this end, we employ a standard non-perturbative procedure based on the Bloch-Feshbach projection operator formalism~\cite{BlochFeshbach1, BlochFeshbach2, Cohen}.

\begin{figure}[b]
\includegraphics[width=1.0 \columnwidth]{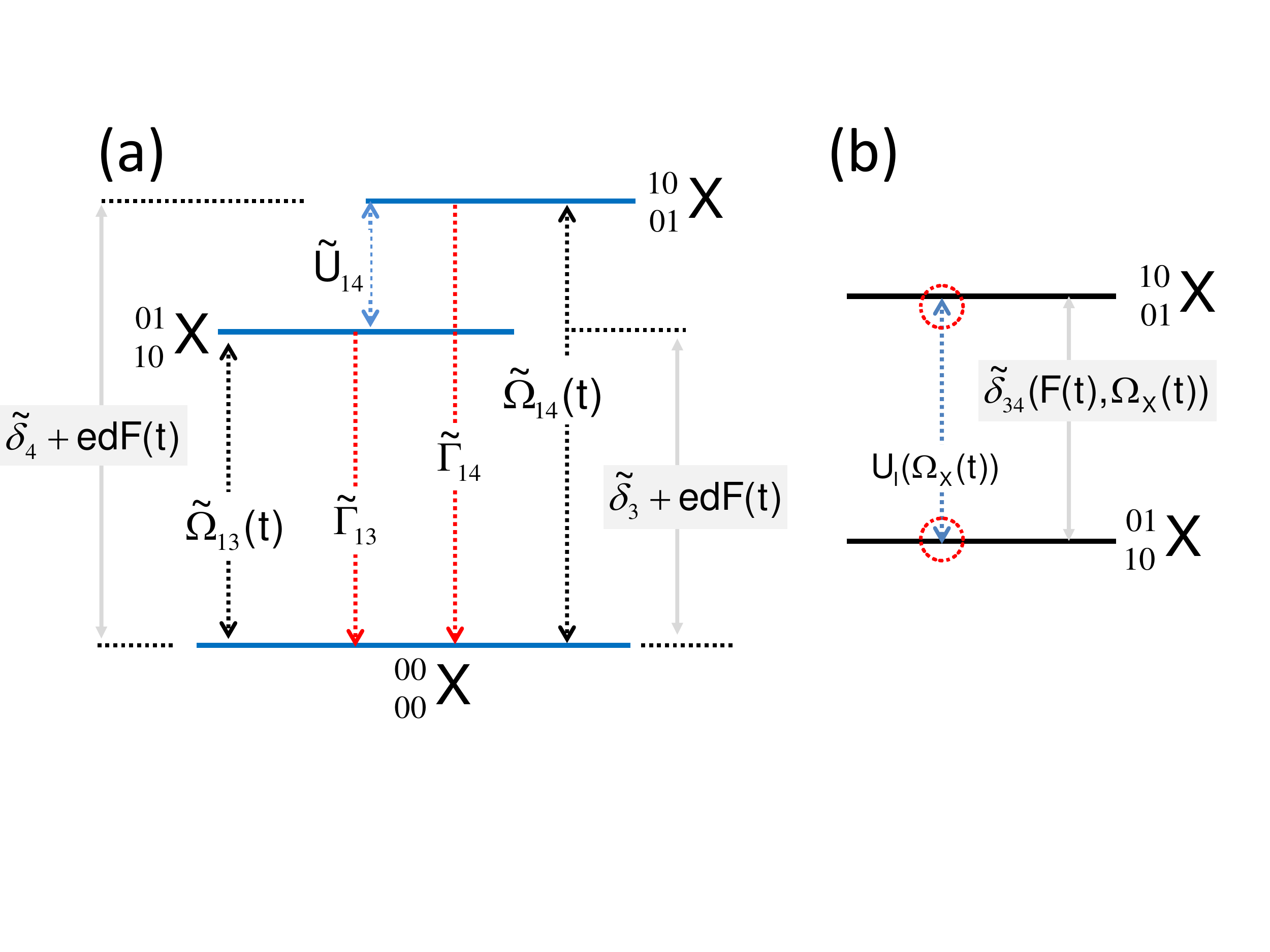}
\caption{(Color online) (a) Effective Hamiltonian level configuration after adiabatically eliminating the direct exciton states. The diagram indicates the effective optical coupling of indirect excitons to the laser field, $\tl{\Omega}_{13}, \tl{\Omega}_{14}$. Dashed red arrows indicate relaxation channels. (b) Effective two-level system (qubit) configuration after projection of the vacuum state onto the indirect exciton subspace. Both, the coupling and level separation are functions of the control parameters, $\Omega_X(t), F(t)$.}
\label{EffectiveHamLevelConfFig}
\end{figure}

To calculate the {\it effective optical couplings} of the spatially indirect excitons to the radiation field, namely $\tl{\Omega}_{13}$ and $\tl{\Omega}_{14}$, we define the target subspace for our projection to be spanned by $\lca{\ket{1}\equiv\ket{_{00}^{00}X},\ket{3}\equiv\ket{_{10}^{01}X},\ket{4}\equiv\ket{_{01}^{10}X}}$, see Fig.\ \ref{EffectiveHamLevelConfFig}(a). The projection procedure yields
\begin{equation}
\tl{\Omega}_{13}= \frac{t_h \Omega_X }{z-\delta_5} + \frac{\left(\Omega_X + \frac{V_F \Omega_X}{z-\delta_5} \right) \left(t_e+\frac{t_h V_F}{z-\delta_5}\right)}{z-\delta_2-\frac{V_F^2}{z-\delta_5}}\, ,
\label{CouplingVAC3}
\end{equation}
\begin{equation}
\tl{\Omega}_{14} = \frac{t_e \Omega_X }{z-\delta_5} + \frac{\left(\Omega_X + \frac{V_F \Omega_X}{z-\delta_5} \right) \left(t_h+\frac{t_e V_F}{z-\delta_5}\right)}{z-\delta_2-\frac{V_F^2}{z-\delta_5}}\, ,
\label{CouplingVAC4}
\end{equation}
where $z=E \pm i\epsilon$ are the complex eigenvalues of $\mcl{H}_\text{eff}$, and
\begin{eqnarray}
\delta_2 &=& E_2-\hbar\omega_L, \nonumber \\
\delta_5 &=& E_5-\hbar\omega_L,
\end{eqnarray}
are the detunings of the spatially direct exciton transitions, $|_{10}^{10}X\rangle$ and $|_{01}^{01}X\rangle$, respectively. As shown in Eqs.\ \ref{CouplingVAC3} and \ref{CouplingVAC4}, the largest contribution to the effective optical coupling arises from processes involving exciton pumping followed by tunneling processes. Similarly, the next leading order contribution involves the combined action of exciton pumping, single charge tunneling and exciton hopping processes. Notice that both couplings are different, a fact that is reflected from the antisymmetric and non-resonant nature of the direct exciton transitions, i.e. $\tl{\Omega}_{13} \rightarrow \tl{\Omega}_{14}$ as $\delta_2 \rightarrow \delta_5$.

On the other hand, the resulting effective coupling between both indirect states is
\begin{equation}
\tl{U}_{34} = \frac{t_e t_h }{z-\delta_5} + \frac{\left(t_e + \frac{t_h V_F}{z-\delta_5}\right) \left(t_h+\frac{t_e V_F}{z-\delta_5}\right)}{z-\delta_2-\frac{V_F^2}{z-\delta_5}}\, ,
\label{Coupling34H3x3}
\end{equation}
where the first term shows that electron-hole cotunneling is the leading process that couples the two indirect excitons, with the second term describing single charge cotunneling and exciton hopping. Notice however, that at this stage of the projection procedure, Eq.\ \ref{Coupling34H3x3} does not reveal the optical Stark shift dependence of the anticrossing features shown in Fig.\ \ref{LevelEigenspectrumFig}, as its effects are still embedded in the matrix elements involving the RWA vacuum state $|_{00}^{00}X\rangle$.
To make  this dependence more explicit, we further reduce the target subspace of the projection procedure to a two-level (qubit) subspace spanned by $\lca{\ket{_{10}^{01}X},\ket{_{01}^{10}X}}$, see Fig.\ \ref{EffectiveHamLevelConfFig}(b). This results in an effective Hamiltonian
\begin{equation}
\mcl{H}_I=\left(
\begin{tabular}{cc}
$\delta_3-\Delta_S + \Delta_h$ & $U_I$  \\
$U_I$ & $\delta_4 + \Delta_S + \Delta_t $\,
\end{tabular}
\right),
\label{HeffI}
\end{equation}
where the diagonal terms contain contributions from the bias detuning $\Delta_S=edF$ and energy shifts given by
\beq
\Delta_h = \frac{t_h^2}{z-\delta_5} +  \frac{\left(t_e+ \frac{t_h V_F}{z-\delta_5}\right)^2}{z-\delta_2-\frac{V_F^2}{z-\delta_5}} + \frac{\tl{\Omega}_{13}^2}{z-\Delta_U} \, ,
\eeq
\beq
\Delta_t = \frac{t_e^2}{z-\delta_5} +  \frac{\left(t_h+ \frac{t_e V_F}{z-\delta_5}\right)^2}{z-\delta_2-\frac{V_F^2}{z-\delta_5}} + \frac{\tl{\Omega}_{14}^2}{z-\Delta_U} \, .
\eeq
On the other hand, the off-diagonal coupling is given by
\begin{equation}
U_I = \tl{U}_{34} + \frac{\tl{\Omega}_{13}\tl{\Omega}_{14}}{z-\Delta_U}\, ,
\label{Coupling34H2x2}
\end{equation}
where
\begin{equation}
\Delta_U = \frac{\Omega_X^2}{z-\delta_5} + \frac{\left(\Omega_X+ \frac{V_F \Omega_X }{z-\delta_5} \right)^2}{z-\delta_2-\frac{V_F^2}{z-\delta_5}}\, .
\label{DeltaOmegaX}
\end{equation}
The leading term $\tl{U}_{34}$ in Eq.\ \ref{Coupling34H2x2} originates from charge cotunneling and exciton hopping, and dominates the behavior of $U_I$ for $0 \leq \Omega_X < 1$ meV, i.e. the effects of the optical coupling are not significant in this regime, see Fig.\ \ref{LevelEigenspectrumFig}(b). However, the second term $\frac{\tl{\Omega}_{13}\tl{\Omega}_{14}}{z-\Delta_U}$ dominates the behavior of $U_I$ for $\Omega_X \geq 1$ meV. This feature enables an all-optical coherent control over the spatially indirect excitonic qubit subspace, starting from a regime in which the dynamics is purely dominated by cotunneling to a regime where the dynamics becomes highly controllable by the bias detuning $F(t)$, and the optical coupling $\Omega_X$.

A distinctive effect of the tunability of the optical coupling $\Omega_X$, is the opening and closing of the indirect excitonic anticrossing gap $\Lambda_I$, achieved by means of the optical Stark shift. After diagonalization of Eq.\ \ref{HeffI} we obtain the corresponding gap equation,
\beq
\Lambda_I = \sqrt{4U_I^2+(2\Delta_S+(\delta_4-\delta_3)+(\Delta_t-\Delta_h))^2}\, .
\label{GapEquation}
\eeq
Equation \ref{GapEquation} defines an energy surface over the control parameters, with minima at the critical values of the bias detuning and optical coupling $\Delta_S^{(c)}=edF_c$ and $\Omega_X^{(c)}$, respectively. Therefore, to find the conditions for which the gap vanishes, we set $\nabla \Lambda_I =0$. This yields
\beq
\Omega_X^{(c)} = \frac{2\sqrt{2}}{t_e-t_h} \sqrt{V_F \left(t_h^2+t_e^2\right)-t_h t_e (\delta_2+\delta_5-2 z)} \, ,
\label{OmegaCritical}
\eeq
\beq
F_c = \frac{1}{2ed}\lpa{(\delta_3-\delta_4)+(\Delta_t-\Delta_h)}\Bigr|_{\Omega_X^{(c)}} \, .
\label{FieldCritical}
\eeq
It is important to remark that in the absence of optical excitation $\Omega_X=0$, a non-vanishing indirect excitonic anticrossing signature, i.e. $\Lambda_I \neq 0$, is conditioned by the coupling $\tl{U}_{34}$ in Eq.\ \ref{Coupling34H3x3}, i.e. only by exciton hopping processes and tunneling. For $V_F=0$, the anticrossing gap emerges only by the action of electron-hole cotunneling processes (see first term in Eq.\ \ref{Coupling34H3x3}), while for $V_F \neq 0$, the gap emerges by the action of either electron or hole tunneling. On the other hand, under the influence of the optical Stark effect $\Omega_X \neq 0$, the vanishing-gap conditions in Eqs.\ \ref{OmegaCritical} and \ref{FieldCritical} are also conditioned by both exciton hopping and single carrier tunneling. For $V_F \neq 0$, the condition is fulfilled with the tunneling of either electron or hole, while for $V_F=0$, the condition is fulfilled only when both electron and hole tunneling are different from zero.

\section{Qubit coherent control via the optical Stark effect}
\label{SectionCoherentControl}

\subsection{Proposed control setup}

To illustrate the controllability of the Hamiltonian in Eq.\ \ref{HeffI}, we interpret the tunability of the indirect excitonic gap in terms of its effect on the temporal evolution of the Bloch vector associated to the state of the indirect-excitonic qubit. To this end, we recast Eq.\ \ref{HeffI} as follows,
\begin{equation}
\mcl{H}_I = \alpha\sigma_0 + \beta\sigma_z + U_{I}\sigma_x \, ,
\label{HeffIPauli}
\end{equation}
where $\sigma_0$ is the $2\times2$ identity matrix, $\sigma_z$, $\sigma_x$ are the Pauli matrices, with corresponding coefficients given by the coupling $U_I$ (defined in Eq.\ \ref{Coupling34H2x2}) and
\beq
\alpha = \lpa{\frac{\delta_3+\delta_4}{2}} + \lpa{\frac{\Delta_t+\Delta_h}{2}} \, ,
\label{alphaPauli}
\eeq
\beq
\beta = \lpa{\frac{\delta_3-\delta_4}{2}} - \lpa{\frac{\Delta_t-\Delta_h}{2}+\Delta_S} \, .
\label{betaPauli}
\eeq
\begin{figure}[h]
\includegraphics[width=0.50 \columnwidth]{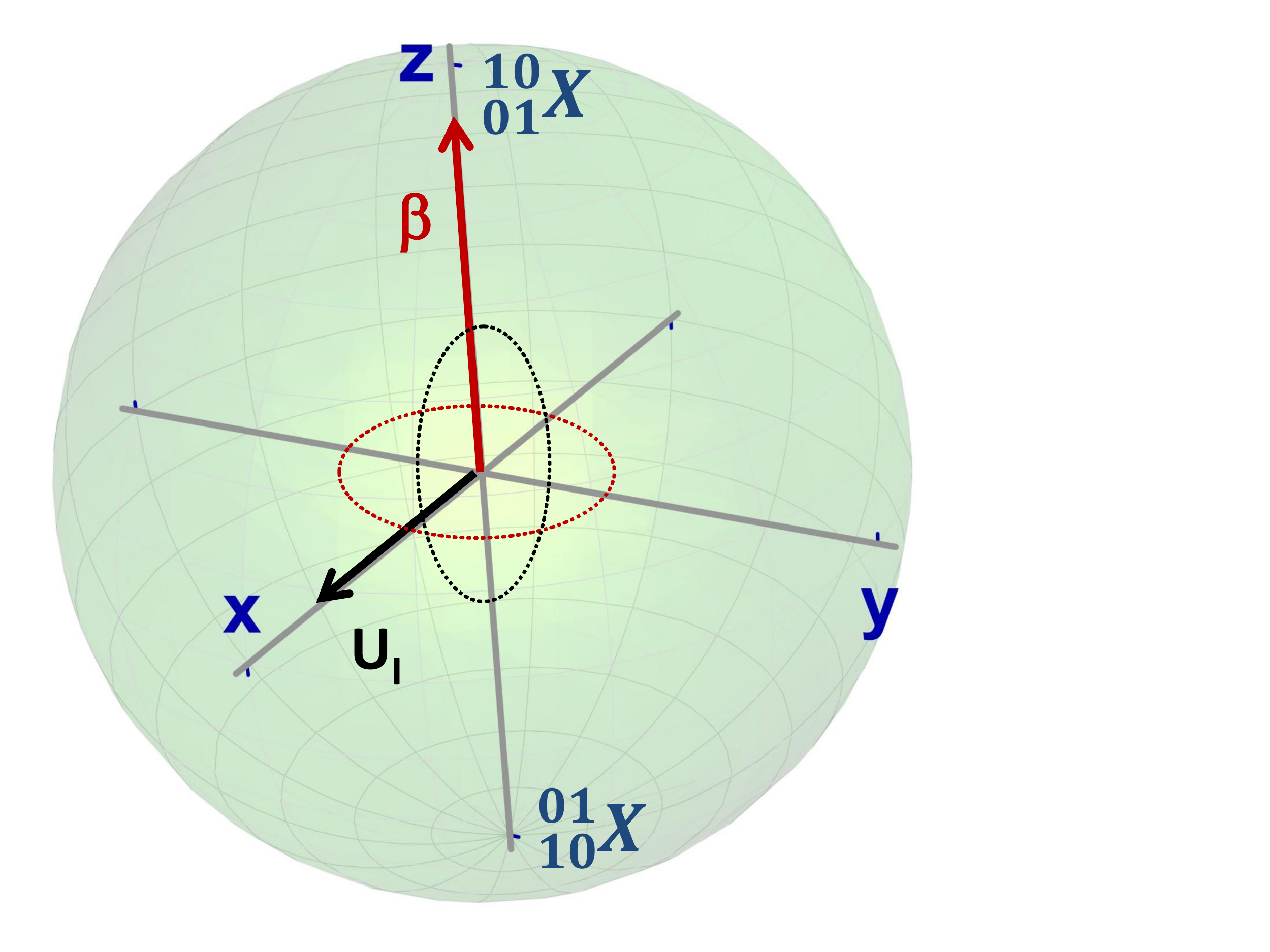}
\caption{(Color online) Bloch sphere geometrical representation of the two-level system ($|_{01}^{10}X\rangle$ and $|_{10}^{01}X\rangle$) and the two rotation axes
($\pmb{\hat{U}_I}$ and $\pmb{\hat{\beta}}$) allowing the implementation controlled rotations of the corresponding Bloch vector.}
\label{BlochSphereFig}
\end{figure}
Assuming fixed values of the variables that depend on the QDM structural and material composition parameters (i.e. single charge confinement energies, single charge tunneling and exciton hopping strengths), time-dependent rotations of the Bloch vector about the $\pmb{\hat{z}}$-axis can be controlled primarily by the value of the bias detuning $\Delta_S$, and by $\Omega_X$-dependent energy shifts $\Delta_h$, $\Delta_t$, as shown in Eq.\ \ref{betaPauli} and Fig.\ \ref{BlochSphereFig}. On the other hand, rotations about the $\pmb{\hat{x}}$-axis are controlled by coupling $U_I$, whose strength is modulated solely by the optical coupling $\Omega_X$. Therefore, a coherent control scheme via Eq.\ \ref{HeffIPauli} allows the implementation of arbitrary rotations over two axes of the Bloch Sphere, opening the possibility to implement universal indirect excitonic qubit operations in QDMs~\cite{NielsenChuang,LossQuantumComp}.

\begin{figure}[h]
\includegraphics[width=1.0 \columnwidth]{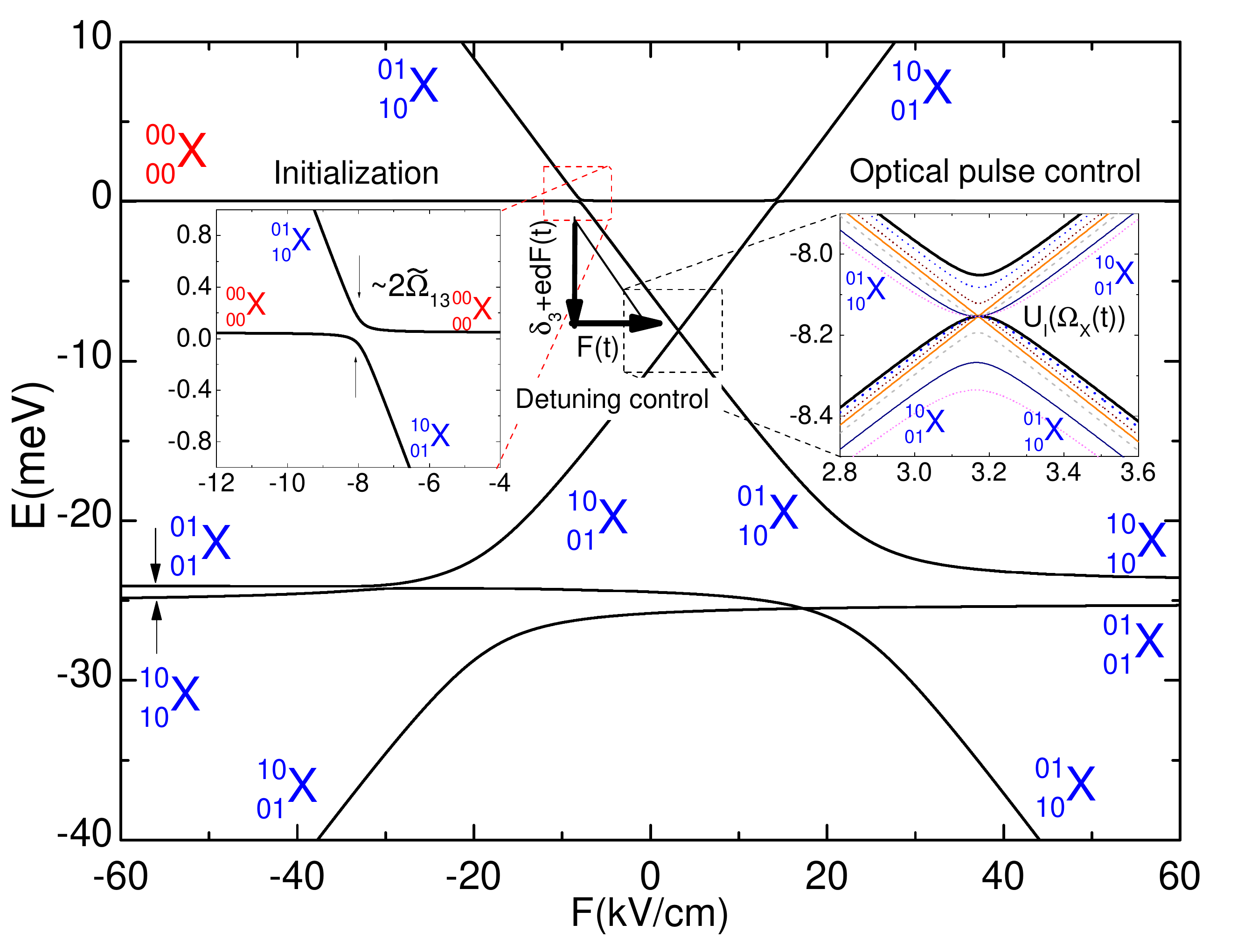}
\caption{(Color online) Level anticrossing spectrum showing schematics of the coherent control scheme. The system is pumped at $\hbar\omega_L =1277$meV with $\Omega_X=0.75$meV. The system is initialized into the state $\ket{_{10}^{01}X}$ at the avoided crossing $\tl{\Omega}_{13}$ at $F=-7.967$kV/cm (dashed red box - left inset). Subsequently, the system is driven by a detuning control bias sweep $\delta_3 +edF(t)$ into the avoided crossing at $F=3.1759$kV/cm (dashed black box). At the avoided crossing the optical coupling is subjected to a temporal Rosen-Zener pulse $\Omega_X(t)$, which controls dynamically the width of the indirect excitonic gap (right inset).}
\label{LACSCoherentControlFig}
\end{figure}

To implement universal coherent operations on the indirect excitonic qubit subspace, we constructed a control scheme based on the combination of a bias time-dependent linear sweep $F(t)$, and a pulsed optical Stark interaction $\Omega_X(t)$. The bias sweep serves two purposes: firstly, it is used to initialize the qubit by selecting a value of the bias detuning such that the radiation field (RWA vacuum $\ket{_{00}^{00}X}$) becomes resonant and coupled (with strength $\tl{\Omega}_{13}$) to one of the indirect excitons $\ket{_{10}^{01}X}$, see red dashed box and inset in Fig.\ \ref{LACSCoherentControlFig}; secondly, the bias sweep controls the indirect exciton detunings $\delta_3 + edF(t)$ and $\delta_4 + edF(t)$ within the qubit manifold. This sweep brings the indirect transitions in and out of resonance from the point of closest approach at the anticrossing mixing $\ket{_{10}^{01}X}$ and $\ket{_{01}^{10}X}$, see Fig.\ \ref{LevelAnticrossingMapFig}(a) and dashed black box in Fig.\ \ref{LACSCoherentControlFig}. Subsequently at the anticrossing, the system is driven with a pulsed optical Stark shift $\Omega_X(t)$, that controls the strength of $U_I$, and consequently the width of the anticrossing gap.

In the present work, we consider a pulsed interaction with a smooth rise profile which controls the strength of the optical Stark effect within the qubit manifold, i.e $U_I$ or equivalently $\Lambda_I$. Among the different choices for the pulse shape profile (e.g. smooth rectangular, Gaussian, hyper Gaussian, etc.)~\cite{Shore}, we employ a {\it Rosen-Zener} hyperbolic tangent profile~\cite{RosenZener, Vitanov}. This choice is made firstly because it conforms well with the assumed non-resonant excitation conditions and energy level structure in our model, and secondly because it produces an efficient population transfer between pairs of anti-crossed states, while minimizing non-resonant population oscillations into states outside the controlled target subspace~\cite{Economou}. The form of our RZ pulse is as follows,
\begin{equation}
\Omega_X(t) = \Omega_X(t_i) + (\Omega_X(t_f)-\Omega_X(t_i))\tanh{\lpa{\frac{t}{\alpha \tau}}}\, ,
\label{RosenZenerPulse}
\end{equation}
where $\tau = t_f - t_i$ defines the time over which the pulsed interaction is non-stationary, and $\alpha \geq \tau^{-1}$ controls the pulse rise. The form of this pulse causes the optical coupling to increase monotonically from an initial value towards a constant value and it has great applicability in the coherent control of multi-level systems approaching the two-state limit, such as the spectrally isolated indirect exciton manifold shown in Fig.\ \ref{EffectiveHamLevelConfFig}(b).

\begin{figure}[t]
\includegraphics[width=1.0 \columnwidth]{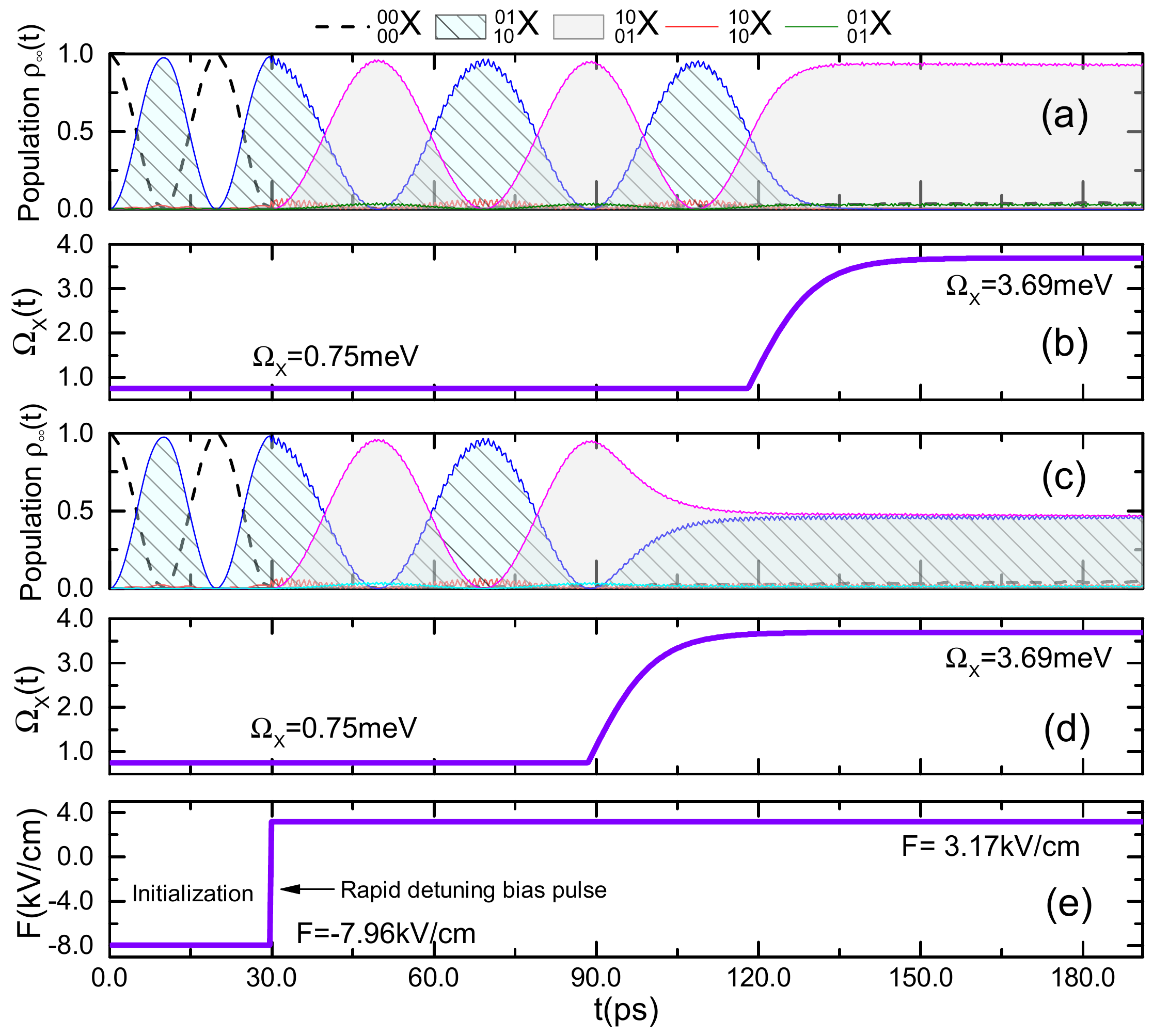}
\caption{(Color online) Exciton population dynamics resulting from the control procedure shown in Fig.\ \ref{LACSCoherentControlFig}. (a and b) The system is initialized in the state $\ket{_{10}^{01}X}$ and brought into the anticrossing mixing $\ket{_{10}^{01}X}$ and $\ket{_{01}^{10}X}$ with a rapid detuning bias pulse shown in (e). After a few Rabi rotations, a pulsed Rosen-Zener optical Stark shift $\Omega_X(t)$, transfers the population into the state $\ket{_{01}^{10}X}$, effectively trapping this state. (c and d) A temporal shift applied to $\Omega_X(t)$ traps the system into an equally weighted superposition of $\ket{_{10}^{01}X}$ and $\ket{_{01}^{10}X}$.}
\label{PopulationDynamicsFig}
\end{figure}

\subsection{Resulting population dynamics}

Figure \ref{PopulationDynamicsFig} shows the exciton population dynamics resulting from applying the control procedure shown in Fig.\ \ref{LACSCoherentControlFig}. The system is initialized with an optical coupling of $\Omega_X=0.75$meV at the anticrossing located at $F=-7.96$kV/cm,  which mixes the RWA vacuum $|_{00}^{00}X\rangle$ and the indirect state $|_{10}^{01}X\rangle$. This anticrossing has a gap $2\tl{\Omega}_{13} \simeq 0.2$meV, such that the Rabi half-period corresponding to a $\pi$ rotation is $\frac{\hbar\pi}{2\tl{\Omega}_{13}} \simeq 10.2$ps. After a $3\pi$ rotation, a sudden detuning bias pulse (see Fig.\ \ref{PopulationDynamicsFig}(e)) traps and brings the spatially indirect exciton into the qubit manifold defined by the avoided crossing at $F=3.1759$kV/cm, which has a gap $2U_I\simeq 0.1$mev. Here, the $\pi$-rotation period corresponding to coherent oscillations of the indirect excitonic qubit basis has a period of $\frac{\hbar\pi}{2 U_I} \simeq 20.3$ps, which can be controlled by the strength of the optical Stark effect. At an elapsed time $t=118.03$ps we apply a RZ pulse with a pulse rise time $\tau = 40.3$ps and rate constant $\alpha=0.2$, which increases the optical coupling from $\Omega_X=0.75$meV up to the critical value $\Omega_X^{(c)} \simeq 3.69$meV; the RZ pulse has the effect of closing the avoided crossing and trapping the exciton populations with a weight determined by the pulse duration $\tau$, and the rise rate constant $\alpha$, see Eq.\ \ref{RosenZenerPulse}. Fig.\ \ref{PopulationDynamicsFig}(a) shows the effect on the population dynamics of the RZ pulse in (b), which traps the exciton $_{01}^{10}X$ with near unity fidelity. On the other hand, Fig.\ \ref{PopulationDynamicsFig}(c) shows the effect of a temporal shift $\delta\tau = 29.6$ps for the same RZ pulse; in this case the system evolves into an equally weighted superposition of $\ket{_{10}^{01}X}$ and $\ket{_{01}^{10}X}$. Clearly, the population dynamics of the indirect excitons corresponds to that of a TLS model, thus corroborating the validity of the analytical expressions obtained by the Bloch-Feshbach projection procedure, and the applicability of the RZ pulse shaping. Note, that there is a small residual (fast oscillating) population of the spatially direct exciton $\ket{_{10}^{10}X}$ (red line) due to the fast dynamics taking place outside the qubit manifold, which originates by the proximity of the electron-tunneling anticrossing; in addition, the indirect exciton manifold at the avoided crossing exhibits resilience against relaxation processes, as shown by the slow rise in the population amplitude of the state $\ket{_{00}^{00}X}$ for $t > 150$ps, see black dashed line in Figs.\ \ref{PopulationDynamicsFig}(a) and (c).

\begin{figure}[htb]
\includegraphics[width=0.50 \columnwidth]{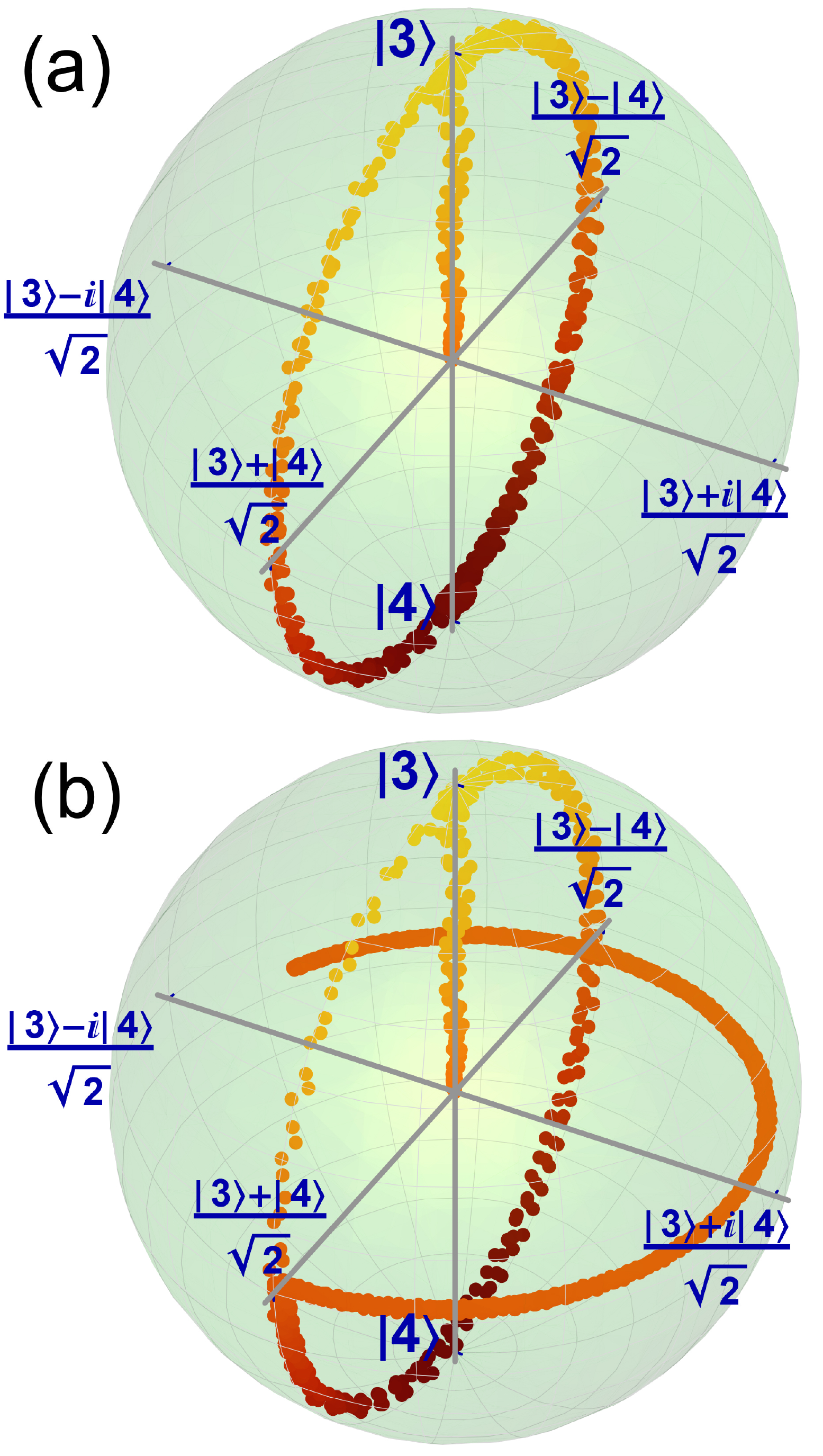}
\caption{(Color online) Bloch sphere representation of the operations resulting from the combined action of the optical RZ and bias $F(t)$ pulses shown in Fig.\ \ref{PopulationDynamicsFig}(b-d-e). (a) Using a rapid detuning bias pulse the Bloch vector is initialized in the state $\ket{3}=\ket{_{01}^{10}X}$, subsequently subjected to a $5\pi$ rotation about the $\pmb{\hat{x}}$-axis. At the end of the RZ pulse the Bloch vector is effectively trapped at the south pole into the state $\ket{4}=\ket{_{10}^{01}X}$. Note that this operation constitutes a concatenation of several Pauli-X quantum gates. (b) A temporal shift of the RZ pulse (see Fig.\ \ref{PopulationDynamicsFig}(d)), causes the Bloch vector to precess about the $\pmb{\hat{z}}$-axis, creating a coherent superposition states of both indirect states; the resulting operation is equivalent to the application of a Hadamard quantum gate operation.}
\label{BlochDynamicsFig}
\end{figure}

To highlight the operational significance of the controlled exciton dynamics in Fig.\ \ref{PopulationDynamicsFig} within the Bloch sphere of the qubit subspace $\lca{\ket{_{10}^{01}X}, \ket{_{01}^{10}X}}$ (see Fig.\ \ref{BlochSphereFig}), we reconstruct the temporal evolution of the Bloch vector via the full numerical solutions for the density matrix dynamics obtained via Eq.\ \ref{Lindblad}. To this end, we parameterize the coordinates of the Bloch vector $\pmb{r}(t)$, such that
\begin{eqnarray}
r_x(t)&=&\rho_{34}(t)+\rho_{43}(t), \nonumber \\
r_y(t)&=&i\lpa{\rho_{34}(t)-\rho_{43}(t)}, \nonumber\\
r_z(t)&=&\rho_{33}(t)-\rho_{44}(t).
\end{eqnarray}

Figure \ref{BlochDynamicsFig} shows the Bloch vector evolution corresponding to the dynamics shown in Fig.\ \ref{PopulationDynamicsFig}. As shown in Fig.\ \ref{BlochDynamicsFig}(a), during the initialization step, as the system leaves the subspace $\lca{\ket{_{00}^{00}X}, \ket{_{10}^{01}X}}$, and enters the qubit subspace, the Bloch vector tip rises from the origin of the Bloch sphere until reaching the north pole at $\ket{3}=\ket{_{10}^{01}X}$. Within the qubit subspace, the Bloch vector precesses  performing several full rotations about the $\pmb{\hat{x}}$-axis under the influence of the interaction $U_I$, see Eq.\ \ref{Coupling34H2x2}; subsequently, under the action of RZ pulse, the system is trapped in the state $\ket{4}=\ket{_{01}^{10}X}$ at the south pole of the Bloch sphere. Interestingly, this operation constitutes a concatenation of several Pauli-X quantum gates, by which the qubit basis states are mapped amongst each other~\cite{NielsenChuang}. On the other hand, Fig.\ \ref{BlochDynamicsFig}(b) shows the effect of applying a temporal shift to the RZ pulse. In this case, the RZ pulse brings the system into the superposition state of both indirect excitons, with the Bloch vector precessing about $\pmb{\hat{z}}$-axis, with a precession frequency dominated by the $\beta$ interaction, see Eq.\ \ref{betaPauli}. Notably, this operation represents the action of the Hadamard quantum gate, by which the qubit basis states are mapped into their symmetric and antisymmetric superpositions~\cite{NielsenChuang}. As shown above, the Bloch vector evolution is fully consistent with the structure of the qubit Hamiltonian in Eq.\ \ref{HeffIPauli}.

\begin{figure}[h]
\includegraphics[width=1.0 \columnwidth]{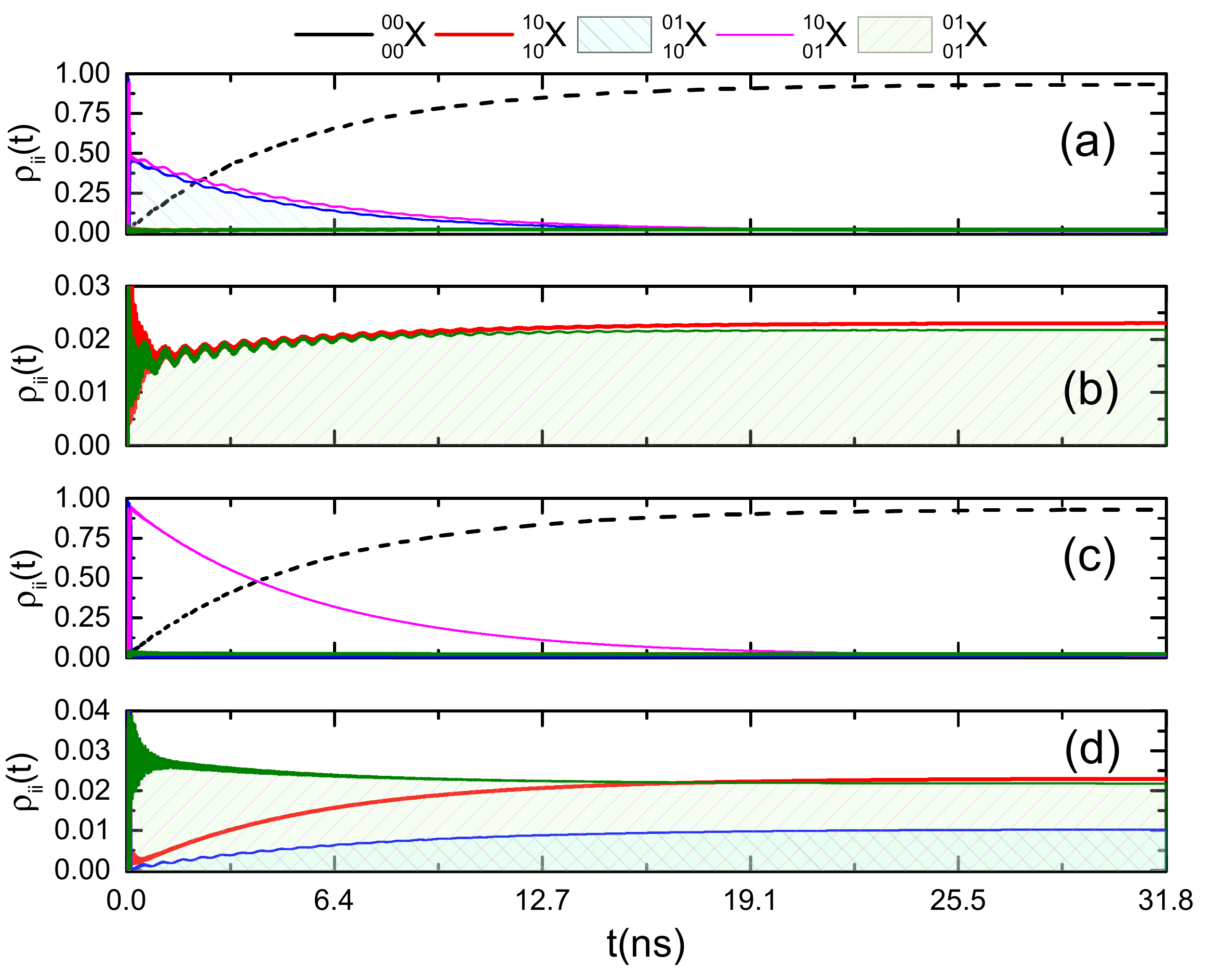}
\caption{(Color online) Long-term relaxation dynamics of the exciton population for times beyond which the RZ pulse reaches its constant critical value $\Omega_X^{(c)}=3.69$meV. (a) Relaxation dynamics of the state $\ket{\Psi_I} \sim \frac{1}{\sqrt{2}} \lpa{\ket{_{10}^{01}X} + \ket{_{01}^{10}X}}$. This state primarily relaxes into the RWA vacuum $\ket{_{00}^{00}X}$, accompanied by a slight filling of the direct exciton states as shown in green and red in (b). (c) Relaxation dynamics of the state $\ket{_{01}^{01}X}$. Besides the RWA vacuum saturation, there is a slight filling of the indirect-direct pair of states connected by electron hopping shown in blue and red in (d).}
\label{LongTermDynamicsFig}
\end{figure}

As discussed in Sec.\ \ref{SectionDynamicStark} (Eqs.\ \ref{CouplingVAC3} and \ref{CouplingVAC4}), indirect excitons possess an effective optical coupling to the radiation field, and consequently a finite lifetime. To highlight the relaxation dynamics within the qubit subspace, Fig.\ \ref{LongTermDynamicsFig} shows the long-term temporal evolution of the exciton population for times beyond which the RZ pulse reaches its constant critical value $\Omega_X^{(c)}\simeq3.69$meV. In all cases, the population of the indirect excitons relaxes to the vacuum state $\ket{_{00}^{00}X}$, while the small residual population of the spatially direct states, $\ket{_{10}^{10}X}$ and $\ket{_{01}^{01}X}$, saturates towards a stationary value. Fig.\ \ref{LongTermDynamicsFig}(a) shows the decay of the superposition state $\ket{\Psi_I}$, with both indirect excitons showing opposite relaxation envelopes; small amplitude oscillations are observed in the population of both direct and indirect states, which reflect the influence of the fast transition dynamics outside the qubit subspace driven by electron tunneling. On the other hand, the relaxation of the trapped state $\ket{_{01}^{01}X}$ shown in Fig.\ \ref{LongTermDynamicsFig}(c) is accompanied by a slight relaxation of the direct state $\ket{_{01}^{01}X}$, and population transfer preferentially into
the states $\ket{_{10}^{10}X}$ and $\ket{_{10}^{01}X}$.

\begin{figure}[t]
\includegraphics[width=0.50 \columnwidth]{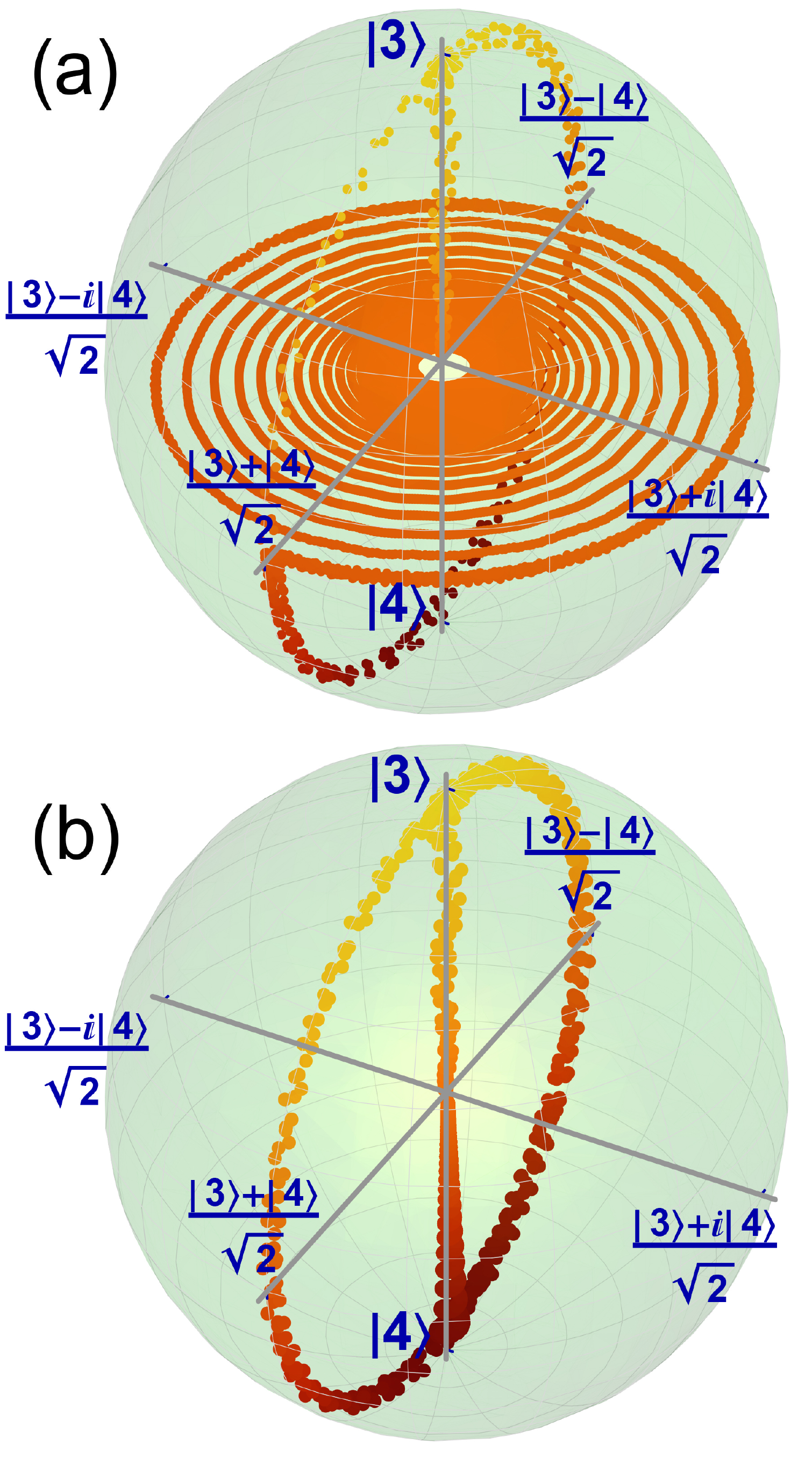}
\caption{(Color online) Bloch sphere representation of the long term exciton relaxation dynamics shown in Fig.\ \ref{LongTermDynamicsFig}. (a) After the initialization and RZ pulsing steps, the Bloch vector for the state $\ket{\Psi_I} \sim \frac{1}{\sqrt{2}} \lpa{\ket{_{10}^{10}X} + \ket{_{01}^{01}X}}$ evolves by precessing about the $z$-axis while its magnitude decreases as it completes several $2\pi$ rotations, eventually collapsing at the origin. (b) The tip of the Bloch vector of the trapped state $\ket{4}=\ket{_{01}^{01}X}$, moves away from the south-pole along the $z$-axis, eventually collapsing at the origin. }
\label{BlochLongTermDynamicsFig}
\end{figure}

Now, Figure \ref{BlochLongTermDynamicsFig} shows the corresponding Bloch sphere representation of the relaxation dynamics presented in Fig.\ \ref{LongTermDynamicsFig}. In Fig.\ \ref{BlochLongTermDynamicsFig}(a), relaxation of the superposition state $\ket{\Psi_I}$ takes place long after the completion of the Hadamard gate. As shown, relaxation induces a decrease in the Bloch vector magnitude as it evolves in time precessing about the $\pmb{\hat{z}}$-axis under the influence of the coupling $\beta$, see Eq.\ \ref{betaPauli}. After an elapsed precession time of $31.8$ns, the Bloch vector eventually collapses into the origin, as the state of the system within the qubit subspace evolves from a pure state into a completely mixed state. On the other hand, Fig.\ \ref{BlochLongTermDynamicsFig}(b) shows the Bloch vector time evolution of the trapped state $\ket{4}=\ket{_{01}^{10}X}$. Here, long after the completion of the Pauli-X gate, the Bloch vector tip moves away from the south-pole along the $z$-axis, eventually collapsing at the origin, as the state of the system leaves the qubit subspace.

\section{Summary and conclusions}
\label{conclusion}

We have characterized the signatures of the optical Stark effect on the spectrum and dynamics of spatially indirect excitons in optically driven quantum dot molecules. By reconstructing the QDM level anticrossing exciton spectrum, we have found an avoided crossing signature between two spatially indirect excitons which persists under strong excitation power broadening effects. Under the influence of the optical Stark effect, the gap of this anticrossing exhibits non-monotonic behavior with a vanishing value that depends primarily on the interplay of charge tunneling and optical excitation.
Using a non-perturbative Bloch-Feshbach projection formalism,
we presented comprehensive analytic results which explain the origin and behavior of the aforementioned signature for different conditions. We have shown that the dynamical opening and closing of the indirect excitonic gap enables a coherent population trapping for indirect excitons, a behavior that is akin to the coherent destruction of tunneling mechanism. We devised a coherent control scheme based on the optical Stark effect, which relies on the variation of the pulse intensity envelope of the pumping laser by means of a RZ hyperbolic tangent pulsed interaction. In particular, within the effective two-level system spanned by the spatially indirect excitons, we defined a qubit manifold whose dynamics is controlled by the combined action of optical RZ and applied electric field sweeps, enabling full control of the Bloch vector across two well defined axes of the Bloch sphere. Our findings pave the way for further research aimed at the design and implementation of indirect-excitonic qubit operations in quantum dot molecules using available ultrafast optical manipulation techniques.

\acknowledgments
This material is based upon work supported by the National Science Foundation under
Grants
No. PHY1306520 (Nuclear Theory program)
and
No. PHY1452635 (Computational Physics program).
J.E.R acknowledges useful discussions with S.E. Ulloa and E. Cota.



\begin{thebibliography}{99}
\bibitem{BiexComment} Biexciton transition energies are far detuned from the single spatially direct and indirect exciton transitions. Therefore, their contribution to the exciton dynamics accounts for higher order perturbative effects which do not qualitatively affect the results of this paper.
\bibitem{parameters} Exciton transition energies and interdot couplings are consistent with Ref.\ [\onlinecite{PawelX, Pawel, UsmanQuantitative, Bester, tunnelings}]. Interdot distance $d$ = $7.5$nm. Interdot couplings $t_{e}, t_h$ = $3.12, 0.13$meV, $V_F=0.11$meV. Exciton bare energies $E_2, E_3, E_4, E_5$ = $1253.1, 1270.7, 1265.9, 1251.9$meV . Interband transition moments $\mu_{T(B)}$ = $6.2e${\AA}, see [\onlinecite{dipoles1,dipoles2,dipoles3}].\@ The exciton-radiation coupling $\Omega_X$ is controlled by the laser intensity envelope. In all cases $\Omega_X\gg\tau_X$, where the exciton relaxation time is given by $\tau_{X}$ = $1.0$ns, see Ref.\ [\onlinecite{lifetimes1,lifetimes2}].
\bibitem{coherent-note-1} In this regime, the population of an exciton $\ket{X}$ at a time $t$ can be arpproximated by $P_X(t)=\vert\langle X\vert U(t)\vert 0\rangle\vert^{2}$, starting with an initial `empty' QDM state $|0\rangle$, with an average population $p_X = (1/\tau_p)\int_{0}^{\tau_p} P_X(t)dt$, where $\tau_p$ stands for a constant amplitude pulse duration.
\end{thebibliography}
\end{document}